\documentclass[review]{elsarticle}
\usepackage{lineno,hyperref}
\modulolinenumbers[5]
\usepackage{color}
\usepackage{xspace}
\makeatletter
\DeclareRobustCommand{\LaTeX}{L\kern-.26em%
        {\sbox\z@ T%
         \vbox to\ht\z@{\hbox{\check@mathfonts
           \fontsize\sf@size\z@
           \math@fontsfalse\selectfont
          A\,}%
         \vss}%
        }%
     \kern-.15em%
    \TeX}
\makeatother
\def\edth{\;\raise1.0pt\hbox{$'$}\hskip-6pt\partial\;}
\def\baredth{\;\overline{\raise1.0pt\hbox{$'$}\hskip-6pt
\partial}\;}
\def\gsim{~\rlap{$>$}{\lower 1.0ex\hbox{$\sim$}}}

\newcommand{\be}{\begin{equation}}
\newcommand{\ba}{\begin{eqnarray}}
\newcommand{\ee}{\end{equation}}
\newcommand{\ea}{\end{eqnarray}}

\newcommand{\abs}[1]{\left\vert#1\right\vert}
\usepackage{graphicx}	
\usepackage{amsmath}	
\usepackage{amssymb}	
\usepackage{hyperref}
\begin{document}
		\title{A new analytical approximation of luminosity distance by optimal HPM-Pad\'e technique}
\tnotetext[mytitlenote]{Fully documented templates are available in the elsarticle package on \href{http://www.ctan.org/tex-archive/macros/latex/contrib/elsarticle}{CTAN}.}

\author[mymainaddress,mysecondaryaddress,mythirdaddress,forzhang]{Bo Yu}


\author[Zhangfirstaddress,forzhang]{Jian-Chen Zhang}
\author[mysecondaryaddress,forzhang]{Tong-Jie Zhang\corref{mycorrespondingauthor}}
\cortext[mycorrespondingauthor]{Corresponding author}
\ead{tjzhang@bnu.edu.cn}
\author[ZhangTingtingfirstaddress]{Tingting Zhang}
\address[mymainaddress]{School of Mathematics  and Big Data, Dezhou University, Dezhou 253023,  China}
\address[mysecondaryaddress]{Institute for Astronomical Science, Dezhou University, Dezhou 253023, China}
\address[mythirdaddress]{Shandong Provincial Key Laboratory of Biophysics, Institute of Biophysics, Dezhou University, Dezhou 253023, China}
\address[Zhangfirstaddress]{School of Computer and Information, Dezhou University, Dezhou 253023, China}
\address[forzhang]{Department of Astronomy, Beijing Normal University, Beijing, 100875, China}
\address[ZhangTingtingfirstaddress]{School of Command and Control Engineering, Army Engineering University , Nanjing 210017, China}
\begin{abstract}
By the use of homotopy perturbation method-Pad\'e (HPM-Pad\'e)  technique, a new analytical approximation of luminosity distance in the flat universe is proposed, which has the advantage of significant improvement for accuracy in approximating luminosity distance over cosmological redshift range within  $0\leq z\leq 2.5$. Then we confront the analytical expression of luminosity distance that is obtained by  our new approach with the observational data, for the purpose of checking whether it works well.  In order  to probe the robustness of the proposed method, we also confront it to supernova type Ia and recent data on the Hubble expansion rate $H(z)$. Markov Chain Monte Carlo (MCMC) code emcee is used in the data fitting. The result indicates that it works fairly well.
\end{abstract}
\begin{keyword}
	\texttt{Theory-distance scale\sep analytical-method\sep Optimal HPM-Pad\'e technique}
\end{keyword}
\maketitle

\section{Introduction}
\label{sec:intr}

Recent astronomical observations  clearly indicate that the universe is currently expanding with an increasing speed, and is spatially flat and  vacuum dominated[1,2]. In order to explain this mysterious 
phenomenon, many cosmological models[3] have been proposed. Since the relation between cosmological distances and redshift depends on the parameters of underlying cosmological models, the accurate and efficient analytical computation of these cosmological distances  becomes an important issue for the comparison of different cosmological models with observation data in modern precision cosmology.

As is well known to all, in the general lambda cold dark matter ($\Lambda$CDM), luminosity distance which is the most important distance from an observational point of view, can only be expressed in the term of  integrals over the cosmological redshift, and computation pressure of the integral of luminosity distance is usual very large. Therefore, for the purpose of avoiding numerical quadrature , several analytical approaches[4,5,6,7,8,9]to approximate luminosity distance in flat universe have been proposed for decades. Among these methods, one of the most widely used approaches is to apply Taylor expansion to approximate luminosity distance[10,11]. Obviously, Taylor polynomial expanding at $z=0$ may have divergence problem caused by cosmological observations that exceed the limits of it. In fact, the supernova data that we obtain now is at least back to $z\approx2.35$ data available[10,11]. Thus several works suggested [11,13,14] that Pad\'e rational polynomial has the ability to approximate luminosity distance, due to its good convergence property in a relatively larger redshift range.

In addition, by solving the differential equation of luminosity distance, Shchig- olev and Yu obtained two formulae  for approximating luminosity distance with smaller error over a relatively small redshift interval based on homotopy perturbation method(HPM)[9](hereafter Shch17)  and optimal homotopy perturbation method (OHPM)[15], respectively. The HPM was first put forward by He[16] to solve nonlinear differential equations, which yields a very accurate solution via one or two iterations. After that various modifications of HPM[17,18,19]were given by various investigators, such as the OHPM coupled with the least squares method[19,20], optimal homotopy  asymptotic method[21], and so forth. In a word, we can obtain more accurate approximations for  luminosity distance over a relatively small redshift interval,based on the use of HPM technique(or  modifications of HPM ). Thus, to reach a compromise between accuracy and redshift convergence interval, the combination of padé approximant  and HPM are therefore adequate candidates to carry out this goal. In fact, homotopy perturbation method-Pad\'e technique(HPM-Pad\'e) has  been recognized as a good one to apply the series solution to improve the accuracy and enlarge the convergence interval[22,23] in the study of nonlinear differential equations.

Therefore, in this paper, we will apply HPM-Pad\'e to obtain a more accurate approximate analytical expression for luminosity distance  in a relatively larger redshift range, based on solving the differential equation of luminosity distance in a flat universe. The rest of this paper is as follows. In Section 2, we briefly review the differential equation of luminosity distance $d_L$ in a spatially flat  universe. The HPM-Pad\'e  rational approximation  of luminosity distance $d_L$ is given in Section 3. In Section 4,  comparison of our rational approximation polynomial for computing luminosity distance is made with the results obtained other existing methods. Then we confront the analytical approximate expression of luminosity distance that was obtained by HPM-Pad\'e technique with the observational data, for the purpose of checking whether it works well. Note that Markov Chain Monte Carlo (MCMC) code emcee[24] is used in the data fitting. Finally, some brief conclusions are given in Section 5.

\section{Differential equation of luminosity distance in a flat  universe}
\label{sec:met:differ}

The general expression of theoretical modulus in a flat universe is defined as follows[1,2]
\begin{equation}\label{eq1}
\mu{(z)}=5\log_{10}{(\frac{d_L}{Mpc})}+ 25
\end{equation}
where $d_L$ is the luminosity distance. In order to verify theoretical calculation, we difine the luminosity distance $d_L$ of SNe Ia in a flat $\Lambda$CDM universe as follows[25]
\begin{equation}\label{eq2}
d_L(z)=\frac{c(1+z)}{H_0} \int_0^z\frac{dt}{\sqrt{\Omega_m(1+t)^3+\Omega_\Lambda}},
\end{equation}
where  $\Omega_m$ and $\Omega_\Lambda$ are the energy densities corresponding to  matter and cosmological constant, respectively: $\Omega_m +\Omega_\Lambda=1 $, $c$ is the speed of light, $z$ is cosmology redshift, and $H_0$ is the Hubble constant.

As mentioned in Shch17, we define $W(z)$ as follows
\begin{equation}\label{eq3}
W(z)=\Omega_m(1+z)^3+\Omega_\Lambda,obviously, W(z)|_{z=0}=1.
\end{equation}
Then  Eq.(2) can be rewritten as:
\begin{equation}\label{eq4}
\frac{d_L(z)H_0}{c(1+z)}= \int_0^z\frac{dt}{\sqrt{W(t)}}.
\end{equation}
For simplicity,we introduce:
\begin{equation}\label{eq5}
1+z=x,u(x)=\frac{d_L(z)H_0}{cx},
\end{equation}
Then, we get
\begin{equation}\label{eq6}
u(x)= \int_0^{x-1}\frac{dt}{\sqrt{W(t)}}.
\end{equation}
By differentiating the Eq.(6), we can obtain
\begin{equation}\label{eq7}
u^{'}(x)=\frac{1}{\sqrt{W(x-1)}}.
\end{equation}
Combining the Eq.(2),(3),(4),(5) and (6), we have
\begin{equation}\label{eq8}
u^{'}(x){|_{x=1}}=1,u(x)|_{x=1}=0.
\end{equation}
According to Eq.(7) and (8),we can derive the Cauchy problem as follows:
\begin{equation}\label{eq9}
u^{''}+\frac{1}{2}W^{'}(x-1){u^{'}}^3=0;u^{'}(x){|_{x=1}}=1,u(x)|_{x=1}=0.
\end{equation}
where the prime is the derivative with respect to $x$, and $W(x-1)|_{x=1}=1$.

\section{ HPM-Pad\'e  rational approximation  of luminosity distance \texorpdfstring{${d_L}$}{Lg}}

\subsection{ Solution of Homotopy perturbation method}

For the homotopy perturbation technique has  already become standard and concise, its  basic idea can be  referred to[16,17]. In order to solve Eq.(9) by homotopy perturbation technique,  we build the homotopy  as follows:
\begin{equation}
{u''}+c_1+p\left[\frac{1}{2}{W'}(x-1){u'}^3-c_1\right]=0,
\end{equation}
Let us assume that the solution of Eq.(10) in the form of a series in $p$ :
\begin{equation}\label{eq11}
u=u_0 +pu_1+p^2u_2+p^3u_3+\dots,
\end{equation}
Substituting Eq.(11) into Eq.(10), and collecting coefficients with the same power of $p$, we get a set of differential equations :
\begin{align}
p^0:& {u_0}^{''}+c_1=0,\label{eq12}\\
p^1:& {u_1}''+\frac{1}{2}W'(x-1){{u_0}'}^3-c_1=0,\label{eq13}\\
&\dots\dots\dots\dots\notag
\end{align}
From Eq.(9), we can obtain the initial conditions as follows:
\begin{align}\label{eq14}
&{u_0}{|_{x=1}}=0, ~{u_0}'{|_{x=1}}=1;\notag\\
&{u_i}{|_{x=1}}=0, ~{u_i}'{|_{x=1}=0};
\end{align}
where $i\geq 1$.

We now can solve the above  differential equations with the initial conditions Eq.(14). Thus, we successively get
\begin{equation}\label{eq15}
u_0=(x-1)-\frac{c_1}{2}(x-1)^2,
\end{equation}
\begin{equation}\label{eq16}
\begin{split}
u_1=&\frac{1}{2}(x-1)+\frac{c_1}{2}(x-1)^2-\frac{1}{2}\int_1^xW(t-1)\\
&\times{\left[1-c_1(t-1)\right]}^2\{1-c_1\left[4(t-1)-3(x-1)\right]\}dt,
\end{split}
\end{equation}
and so on. By setting $p=1$~in  Eq.(11),we get  the approximate analytical  expression of luminosity distance

\begin{equation}\label{eq17}
\widetilde{d}_L(z)=\frac{c(1+z)}{H_0}\{z+\alpha_1z^2+\alpha_3z^3+\dots+
+\alpha_kz^k+\dots\}
\end{equation}
where $c$ is the speed of light, $H_0$ is the Hubble constant, $z$ is redshift, and $ \alpha_k(k \geq 1)$ are functions of the unknown constant $c_1$.

\subsection{Pad\'e approximate technique to the series solution of differential equation for luminosity distance \texorpdfstring{${d_L}$}{Lg}}

Pad\'e approximate technique is the best  approximation of a function $f(z)$ by a rational polynomial of a given order($m$,$n$)[26]
\begin{equation}\label{eq18}
P_{mn}(z)=\frac{a_0+a_1z+\dots+a_mz^m}{1+b_1z+\dots+b_nz^n}
\end{equation}
In order to determine coefficients of Pad\'e approximation of order($m$,$n$), let us assume $f(z)$ can be expanded in the form of a power series in $z$ as follows:
\begin{equation}\label{eq19}
f(z)=\sum\limits_{k=0}^{m+n+1}{d_kz^k}
\end{equation}
Generally, $f(z)$ is expanded in Taylor series about at the point $z=a$. In this study, as mentioned above in Section 2,  the  approximate analytical  expression of luminosity distance $\widetilde{d}_L$ obviously has similar form  of a power series in $z$. For convenience, let  us rewrite the approximate analytical  expression of luminosity distance(Eq.(17)) $\widetilde{d}_L$  as 
\begin{equation}\label{eq20}
\widetilde{d}_L(z)=\sum\limits_{k=0}^{m+n+1}{d_kz^k}
\end{equation}

So,we obtain
\begin{equation}\label{eq21}
\widetilde{d}_L(z)-\frac{a_0+a_1z+\dots+a_mz^m}{1+b_1z+\dots+b_nz^n}=O(Z^{n+m+1})
\end{equation}

Then the Eq.(21) can be  written  out as
\begin{equation}\label{eq22}
\begin{split}
&d_{m+1}+d_{m}b_1+\dots+d_{m-n+1}b_n=0\\
&d_{m+2}+d_{m+1}b_1+\dots+d_{m-n+2}b_n=0\\
&\vdots \\
&d_{m+n}+d_{m+n}b_1+\dots+d_{m}b_n=0,
\end{split}
\end{equation}
\begin{equation}\label{eq23}
\begin{split}
&a_0=d_0\\
&a_1=d_1+d_0b_1\\
&a_2=d_2+d_1b_1+d_0b_2\\
&\vdots \\
&a_{m}=d_m+d_{m-1}b_1+\dots+d_0b_m,
\end{split}
\end{equation}
From Eq.(22),we can get the $b_i(1\leq i \leq n)$. By the known values of $b_1,b_2,\dots,b_n$, we can obtain the values of  $a_0,a_1,\dots,a_m$ from the  Eq.(23).Then we obtain the approximate analytical  expression of luminosity distance as follows:
\begin{equation}\label{eq24}
\widetilde{d}_L(z)=\frac{c}{H_0}\frac{a_0+a_1z+\dots+a_mz^m}{1+b_1z+\dots+b_nz^n}
\end{equation}
where $c$ is the speed of light, $H_0$ is the Hubble constant, $z$ is redshift, and $b_i(1\leq i \leq n)$ and $a_k(0\leq k \leq m)$ are functions of the unknown constant $c_1$.

The relative error of the approximation for the luminosity distance is given by[5]
\begin{equation}\label{eq25}
\bigtriangleup E=\abs{\frac{\widetilde{d}_L-d_L^{num}}{d_L^{num}}}
\end{equation}
where  $d_L^{num}$  and $\widetilde{d}_L$ stand for the values of luminosity distance calculated from the numerical method  and our HPM-Pad\'e approximate expression, respectively.

Many methods, such as the least square method[19,20] and the collocation method [27], can be used to  optimally determine the unknown constant $c_1$. As similar in our previous work[15], by minimizing the relative error, we  optimally determine unknown constant $c_1$, which yields the following algebraic
equation
\begin{equation}\label{eq26} 
\frac{\partial\bigtriangleup E}{\partial C_1}=0
\end{equation}

Significantly, it is very important to choose the orders ($m,n$) of the rational polynomial for  luminosity distance, which may lead to divergence problems and bias its corresponding numerical results. From one hand, the uncertainties of the free coefficients ($b_i(1\leq i \leq n)$and $a_k(0\leq k \leq m)$) increases when the orders are too high and the number of  coefficients is too many; on the other hand, the accuracy of the  rational polynomial in approximating for  luminosity distance will be small when the number of  coefficients is too little. Given this fact, in order to make sure the rational approximations to be convergent, it is essential to consider a small number of  free coefficients. Zhou choosed a moderate order($2,2$) in the work of Pad\'e parameterization of the luminosity distance [13]. S. Capozziello pointed out that the order($2,1$) of Pad\'e approximation for the luminosity distance (Hereafter P(2,1)) is better way to explain high-low cosmological redshift data[11].To our knowledge, the best way to this issue on orders  choice of rational polynomial is to analysize
the above mentioned ones with different orders.

\section{Performance of the HPM-Pad\'e  rational approximation }
\label{sec:met:performance}

In this section, the performance of   HPM-Pad\'e  rational approximation in Sect.3 is assessed, which includes two steps. Firstly, comparison of our proposed approach for computing $\Lambda$CDM model luminosity distance is made with the results obtained by other methods. Then we confront the analytical expression of luminosity distance that was obtained by the HPM-Pad\'e technique with the observational data, for the purpose of checking whether it works well. 

\subsection{HPM-Pad\'e  approximation versus other methods for the \texorpdfstring{$\Lambda CDM$}{Lg} model }
\label{sec:4.1}

To help visualize this goal, we  will give a qualitative representation of the improvements for accuracy in approximating luminosity distance, which is obtained by  performing  HPM-Pad\'e technique.

\begin{figure}[tbp]
	\centering
	\includegraphics[width = 1.1\textwidth]{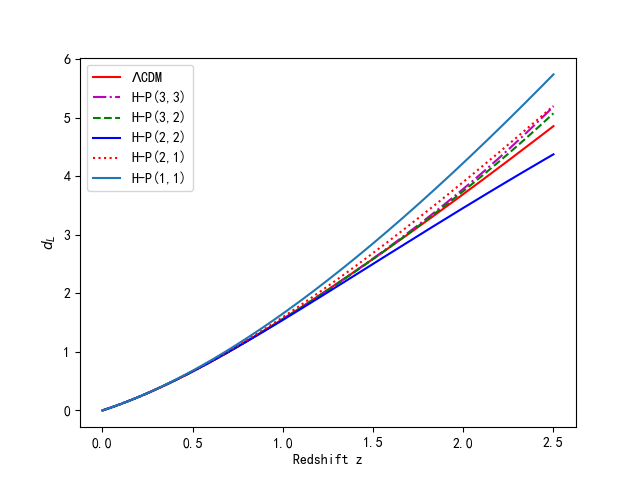}
	\caption{ The  curves of luminosity distance  $d_L$ for the $\Lambda$CDM model when $\Omega_m=0.28$ and the comparison with the numerical behavior of HPM-Pad\'e approximation with  different  orders over the range of redshift  $0< z\leq 2.5$.}
\end{figure}

\begin{figure}[tbp]
	\centering
	\includegraphics[width = 1.1\textwidth]{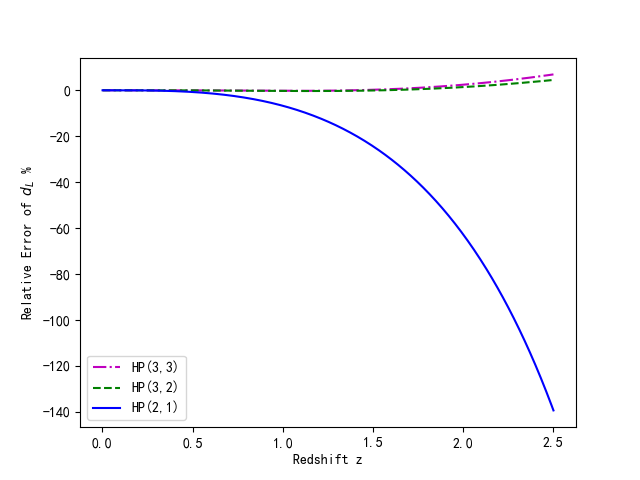}
	\caption{When  $\Omega_m=0.28$, relative error percentages of rational approximations for $d_L$($\bigtriangleup E$) in the $\Lambda$CDM model with third order H-P(2,1), fifth order H-P(3,2) and sixth order H-P(3,3), when  $\Omega_m=0.28$. }
\end{figure}

\begin{figure}[tbp]
	\centering
	\includegraphics[width = 1.1\textwidth]{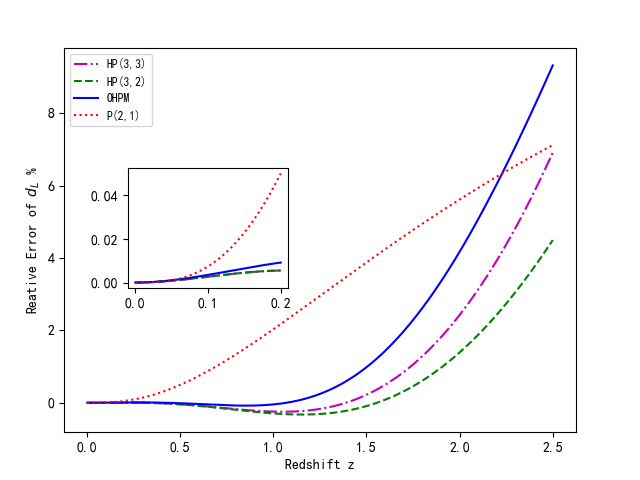}
	\caption{For the fixed  $\Omega_m=0.28$, relative error percentages of approximate for $d_L$($\bigtriangleup E$) in  the $\Lambda$CDM model obtained by different approximation techniques.The relative error percentages of $d_L$ over range   $0\leq z\leq 0.2$ is amplified,which is shown in the inset.}
\end{figure}

\begin{figure}[tbp]
	\centering
	\includegraphics[width = 1.1\textwidth]{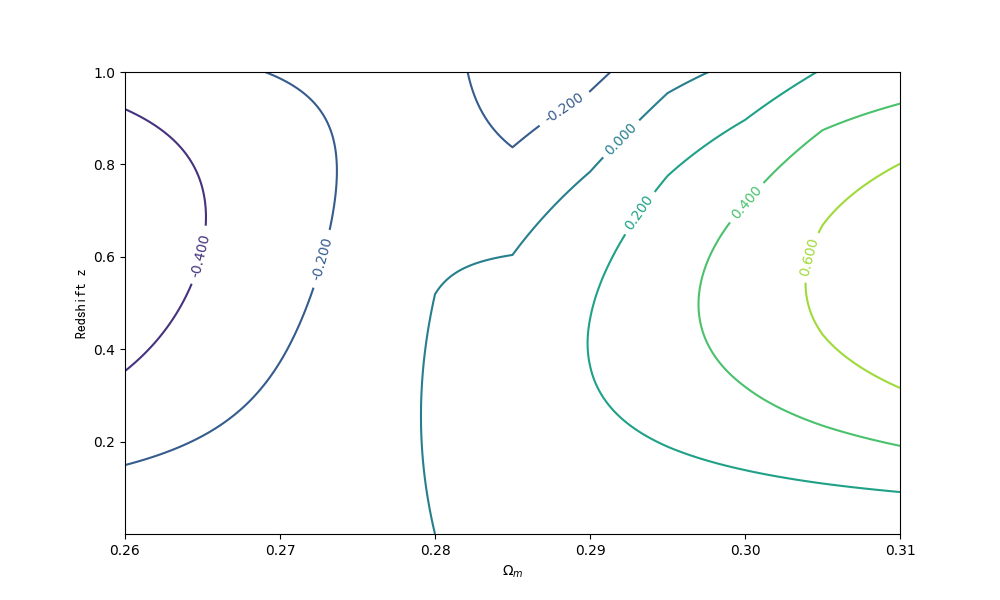}
	\caption{The contour for the distribution of relative error percentages of H-P(3,2) for $d_L$($\bigtriangleup E$) in the $\Lambda$CDM model by  use of HPM-Pad\'e approximate  to  corresponding to $\Omega_m $ within  $0.26\leq \Omega_m\leq 0.31$ and over redshift interval $0\leq z \leq 1$. The  global error is dominated by the``peaks''and``pits'' in the region.}
\end{figure}

\begin{figure}[tbp]
	\centering
	\includegraphics[width=1.1\textwidth]{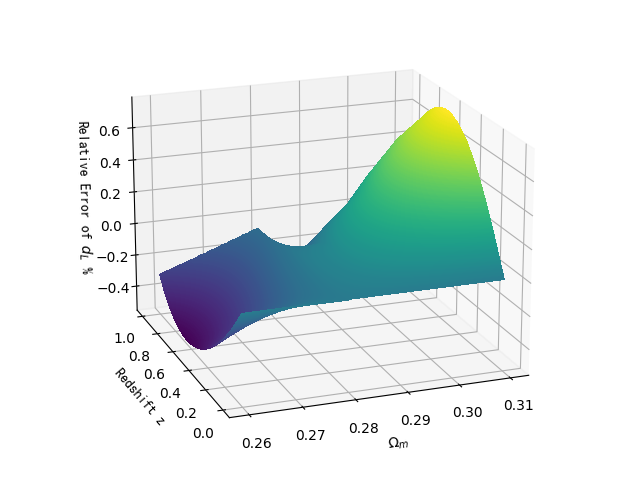}
	\caption{The global relative error surface  of approximation H-P(3,2) for $d_L$($\bigtriangleup E$) in the $\Lambda$CDM model corresponding to $\Omega_m $ over the span $0.26\leq \Omega_m\leq 0.31$. When the variation of $\Omega_m$ from 0.26 to 0.31, the relative error  increases. When the variation of cosmological redshift $z$ from 0 to 1,the relative error first increases and then decreases within the range $0.28\leq \Omega_m\leq 0.31$, and the relative error first decreases and then increases  within the range $0.26\leq \Omega_m\textless 0.28$.}
\end{figure}

For the elucidative purposes, in Fig.1, we plot analytical curves of luminosity distance  $d_L$ for the $\Lambda$CDM model and the comparison with the numerical behavior of HPM-Pad\'e approximation with  different orders over the range of redshift  $0\leq z\leq 2.5$. From the  Fig.1, rational  HPM-Pad\'e polynomial of third order H-P(2,1), fifth order H-P(3,2) and sixth order H-P(3,3) have been selected according to their good behaviors over the range of redshift  $0\leq  z\leq 2.5$. Fig. 2 shows the relative error percentages of  $d_L$($\bigtriangleup E$)for third order H-P(2,1), fifth order H-P(3,2) and sixth order H-P(3,3), which indicates that fifth order H-P(3,2) and sixth order H-P(3,3) behave  significantly well in approximating $\Lambda$CDM model luminosity distance.

In Fig. 3, the comparison of fifth order H-P(3,2) and sixth order H-P(3,3) for computing $\Lambda$CDM model luminosity distance is made with the results obtained by some existing methods  such as OHPM and P(2,1).The HPM-Pad\'e  approximation is superior to other methods, because the relative error percentage  $\bigtriangleup E\approx 4\%$ is reached at $z=2.437$ for HPM-Pad\'e  approximation  with order(3,2), at $z=1.959$ for optimal homotopy perturbation method, and at $z=1.540$ for Pad\'e  approximation with order(2,1); the  relative error percentages  $\bigtriangleup E$ at $z=0.2$ are 0.0053\% for HPM-Pad\'e  approximation  with order(3,2)(or  order(3,3)), 0.0099\% for optimal homotopy perturbation method, and 0.0498\% for Pad\'e  approximation with order(2,1), respectively. Fig.4 shows that  the relative error percentages range of H-P(3,2) for $d_L$ in $\Lambda$CDM model is from -0.4\% to 0.6\% for any cosmological redshift within   $0\leq z\leq 1$ over the $\Omega_m$ span  $0.26\leq \Omega_m\leq 0.31$. Seen from  Fig.5, we can get that  the relative error  increases  when the variation of $\Omega_m$ from 0.26 to 0.31; the relative error first increases and then decreases within the range $0.28\leq \Omega_m\leq 0.31$, and the relative error first decreases and then increases  within the range $0.26\leq \Omega_m\textless 0.28$.

Obviously, the above figures indicate that, provided we are given data over the relatively larger cosmological redshift interval, HPM-Pad\'e technique would be better way to fit the observed luminosity distances by means of a rational polynomial, for the purpose of getting a more accurate  approximate function. Given this fact,we can obtain a better bounds on the parameters of rational HPM-Pad\'e  approximation polynomial, which is exactly what we expect.

\subsection{Experimental analysis of Type Ia supernova and OHD data with HPM-Pad\'e approximation }
\label{sec:4.2}

Based on the  HPM-Pad\'e  approximation polynomial of luminosity distance, we can obtain the distance modulus approximation as a function of $H_0$, $z$ and $c_1$:
\begin{equation}\label{eq27}
\widetilde{\mu}{(z)}=5\log_{10}{(\frac{\widetilde{d}_L(z)}{Mpc})}+ 25
\end{equation}

Then we confront it with SNIa data to constrain the  approximate analytical  expression of luminosity distance(Eq.(24)), by use of relation between SNIa data and luminosity distance. Here, we consider the ``Pantheon Sample"  consisting of 1048 data points within the redshift range $0.01\leq z \leq 2.3$[12] in terms of the distance modulus $\mu_{obs}{(z_i)}$. By [13], the $\chi^2$ from ``Pantheon Sample" 1048 SNIa data can be defined as

\begin{equation}\label{eq28}
\chi^2_{SN}= \sum\limits_{i=1}^N \frac{(\mu_{obs}{(z_i)}-\widetilde{\mu}{(z_i)})^2}{\delta^2(z_i)}
\end{equation}
where $N=1048$, and the $\delta(z_i)$ is corresponding error of observed distance modulus $\mu_{obs}{(z_i)}$ at $z_i$. Correspondingly, a reduced merit function $\chi^2_{red}$ becomes:

\begin{equation}\label{eq29}
\chi^2_{red}= \frac{\chi^2_{SN}}{NF}
\end{equation}
where the number of degrees of freedom $NF$ is equal to $N-k$, $N$ is the number of SNIa data, and $k$ is the number of parameters of $\widetilde{\mu}{(z)}$. The Akaike information criterion (AIC) [28] is defined by

\begin{equation}\label{eq30}
AIC= 2k-2\ln (L_{max})
\end{equation}
where $L_{max}$ is the maximum likelihood function.

In order to get the best fits, Markov Chain Monte Carlo (MCMC) code emcee[24] was applied on SNIa likelihood. Table 1 reports the numerical results for two parameters of  HPM-Pad\'e approximations of luminosity distance. From Table 1, we can see that the constrains on two parameters of rational HPM-Pad\'e  approximation polynomial are significantly tightened. Fig.6 shows the best fit for distance modulus(HPM-Pad\'e approximation of order(3,2)).By confronting the HPM-Pad\'e approximation of luminosity distance Eq.(24) with the cosmological observational
data, we clearly see that it works fairly well.

As mentioned in[29,30], to probe the robustness of the proposed method, we also confront it to the combination of SNIa data and  recent data on the Hubble expansion rate $H(z)$,which includes 31 $H(z)$ measurements that are determined by using the cosmic chronometric technique(see Table 1 in [31]). In fact, we can insert Eq.(24) into

\begin{equation}\label{eq31}
H(z)=\left[ {\frac{d}{dz}\left(\frac{d_L(z)}{1+z}\right)}\right]^{-1} 
\end{equation}
Table 2 reports the numerical results for two parameters of  HPM-Pad\'e approximations  with order(3,2)and order(3,3). From Table 2, we can see that the best-fitting values are obtained  by HPM-Pad\'e approximation of order(3,3).  The marginal distribution function in 1D parameter space, and 1$\delta$, and 2$\delta$  contour in the 2D parameter spaces for the best fitting from  SNIa data, combination of SNIa data and OHD data are shown in Fig.7. According to  Fig.7, we can get that the constrains on two parameters of rational HPM-Pad\'e  approximation polynomial  from  SNIa data are in good agreement with ones from combination of SNIa data and OHD data. From Table 1 and Table 2, by adding OHD data, the constrains on two parameters are more tightened.   

\begin{table}
	\centering
	\caption{MCMC results at the 68.3\% (95.4\%) C.L for HPM-Pad\'e approximations of luminosity distance  from SNIa data. The corresponding  AIC and $\chi^2_{red}$ are also shown.}.
	\begin{tabular}{|l|c|c|c|r|}
		\hline
		Model & $H_0$ & $c_1$ & AIC & $\chi^2_{red}$ \\
		\hline
		HP(3,2) &$72.044_{-0.076(-0.149)}^{+0.075(+0.154)}$ & $0.4171_{-0.019(-0.038)}^{+0.021(+0.041)}$ & 1036.494 & 0.9890 \\
		\hline
		HP(3,3) &$72.039_{-0.074(-0.143)}^{+0.076(+0.151)}$ & $0.415_{-0.017(-0.034)}^{+0.019(+0.37)}$ & 1036.60 & 0.9891 \\
		\hline
	\end{tabular}
\end{table}

\begin{table}
	\centering
	\caption{MCMC results at the 68.3\% (95.4\%) C.L for HPM-Pad\'e approximations with order(3,2)and order(3,3) from combination of SNIa adta and OHD data. The corresponding  AIC and $\chi^2_{red}$ are also shown. }.\label{tab 2}
	\begin{tabular}{|l|c|c|c|r|}
		\hline
		Model & $H_0$ & $c_1$ & AIC & $\chi^2_{red}$ \\
		\hline
		HP(3,2) &$72.060_{-0.071(-0.141)}^{+0.073(+0.144)}$ & $0.431_{-0.015(-0.027)}^{+0.015(+0.030)}$ & 1455.07261 & 0.97742 \\
		\hline
		HP(3,3) &$72.050_{-0.071(-0.141)}^{+0.072(+0.142)}$ & $0.424_{-0.009(-0.018)}^{+0.0102(+0.021)}$ &1455.07260 & 0.97727 \\
		\hline
	\end{tabular}
\end{table}

\begin{figure}[tbp]
	\centering
	\includegraphics[width=1.1\textwidth]{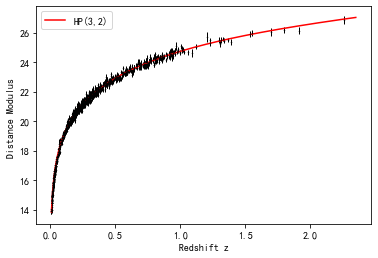}
	\caption{Hubble diagram for Pantheon Sample. The best fit for distance modulus approximation as represented by Eq.(27)  with order (3,2) is  represented the red solid line.}
\end{figure}

\begin{figure}[tbp]
	\centering
	\includegraphics[width=1.1\textwidth]{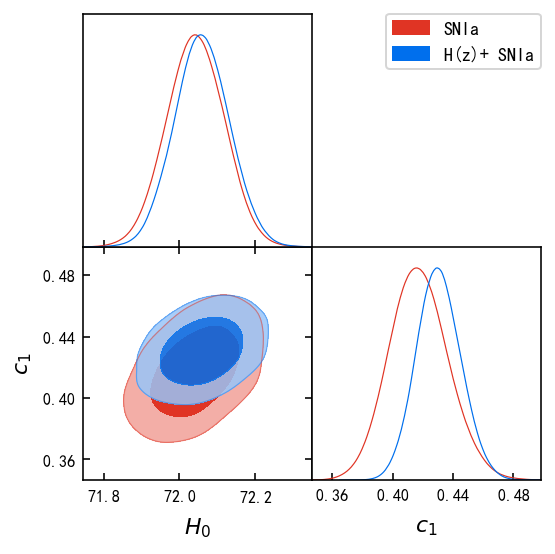}
	\caption{The  contour for 1$\delta$, and 2$\delta$  in the 2D parameter spaces, and  the marginal distribution function in 1D parameter space. The deep shaded region and light shaded region  for each contour stand for 68  percnet confidence level and 95  percent confidence level,respectively.} 
\end{figure}

\section{Conclusion and discussion}
\label{sec:conclusion}

In this paper, based on use of homotopy perturbation method-Pad\'e (HPM-Pad\'e)  technique, a new analytical approximation of luminosity distance in the flat universe is proposed. The  numerical results clearly indicate that HPM-Pad\'e technique has obvious advantages in improving accuracy of approximating luminosity distance over relative larger cosmological redshift range interval. It is worthy noting that the choice of the orders ($m,n$) for luminosity distance rational polynomial is very important, which may yield divergence problems and bias its true numerical results. For the elucidative purposes, we make the comparison of rational approximations with different orders for computing $\Lambda$CDM model luminosity distance in Figure 1. Figure 3 indicates that the HPM-Pad\'e  approximation is superior to other methods. As an example mentioned above in section 4, the relative error percentage  $\bigtriangleup E\approx 4\%$ is reached at $z=2.437$ for HPM-Pad\'e  approximation, at $z=1.959$ for optimal homotopy perturbation method, and at $z=1.540$ for Pad\'e  approximation, respectively. It means that we can get a larger cosmological redshift range of convergence by HPM-Pad\'e  approximation.

According to [29] and [30], to probe the robustness of the proposed method, we  confront it to SNIa data, combination of SNIa data and OHD data, respectively. The results indicate  the rational HPM-Pad\'e  approximation polynomial  is robust. Furthermore, the proposed  analytical approximation of luminosity distance in the flat  $\Lambda$CDM,  has the advantage of significant improvement for accuracy in approximating luminosity distance over cosmological redshift range within  $0\leq z\leq 2.5$. In other words, the proposed  analytical approximation of luminosity distance in the flat  $\Lambda$CDM can be trusted up to $z \sim 2.5$, which can avoid a breakdown in the validity of the approximation for luminosity distance over cosmological redshift range within  $0\leq z\leq 2.5$  that may be misinterpreted as a (phantom) deviation from $\Lambda$CDM.
Due to its model-dependent, rational HPM-Pad\'e  approximation polynomial  of luminosity distance for $\Lambda$CDM model can not be used to model different dark energy behaviors for different range.[32]. To our knowledge, for methodological reasons, we cannot apply HPM-Pad\'e technique to other cosmological models. Take the $\omega$CDM model as  example,  we will obtain a series of fractional powers of redshift $z$ during the process of solving differential equation of luminosity distance. As a result, the rational analytic expression of luminosity distance cannot be obtained by using  Eq.(21), which causes  this technique to be invalid.  To conclude, we have put forward and investigated here HPM-Pad\'e technique in approximating luminosity distance. Moreover, HPM-Pad\'e technique can be used in other fields of precision cosmology, due to its good properties.

\section*{Acknowledgments}
We thank Anonymous Referees for their valuable comments for revising and improving earlier draft of our manuscript. We are grateful to Prof. Jin-Yu He for his kind help.We
are grateful to Kang Jiao for useful discussions. This work was supported by National Science Foundation of China (Grants No. 11573006,11929301), and National Key R\&D Program of China (2017YFA0402600).
\bibliographystyle{elsarticle-num}

\end{document}